\documentclass[12pt]{article}
\usepackage{amsmath}
\usepackage{graphics}
\usepackage{latexsym}

\usepackage [ansinew]{inputenc}
\usepackage {hyperref}
\begin{document}

\vskip 2cm
\title{Open String Fluctuations in AdS with and without
Torsion}
\author{\\
A.L. Larsen${}^{|}$
and
M.A. Lomholt${}^{*}$}
\maketitle
\noindent
{\em Physics Department, University of Southern Denmark,
Campusvej 55, 5230 Odense M,
Denmark}

\vskip 8cm
\noindent
$^{|}$Electronic address: all@fysik.sdu.dk\\
$^{*}$Electronic address: michael@lomholt.name
\newpage
\begin{abstract}
\baselineskip=1.5em
\hspace*{-6mm} The equations of motion and boundary conditions for
the fluctuations around a classical open string, in a curved
space-time with
torsion, are considered in compact and world-sheet covariant form.
The rigidly rotating open strings in Anti de Sitter space with
and without
torsion  are investigated in detail. By carefully analyzing the
tangential fluctuations at the boundary, we show explicitly that
the physical fluctuations
(which at the boundary are combinations of normal and
tangential
fluctuations) are finite, even though the world-sheet
is singular there.
The divergent 2-curvature thus seems less dangerous than expected, in these
cases.

The general formalism can be straightforwardly used also to study the
(bosonic part of
the)  fluctuations around
the closed  strings, recently considered in connection with the
AdS/CFT duality, on AdS$_5\times$S$^5$
and AdS$_3\times$S$^3\times$T$^4$.
\end{abstract}

\newpage
\newpage
\section{Introduction}
\setcounter{equation}{0}
Semi-classical quantization of strings in Anti de Sitter space leads to the
result that
the energy $E$ scales with a quantum number $N$, $E\sim N$ (for large $N$).
This
result, which is independent of the dimensionality of
Anti de Sitter space, was
originally obtained almost a decade ago by considering fluctuations around
the string center
of mass
\cite{sanchez} (with $N$ being a combination of spin angular momentum and
oscillation
number), by considering circular strings \cite{de vega} (with $N$ being the
oscillation
number), and by considering rigidly rotating strings \cite{inigo} (with $N$
being the  spin
angular momentum).

This result has recently received a lot of attention in connection with the
conjectured
duality \cite{maldacena,gubser,witten} between super string theory on ${\rm
AdS_5}\times {\rm S^5}$ and
${\cal N}=4$ SU(N)
super Yang-Mills theory in Minkowski space. In the case of rigidly rotating
strings, it was
noticed \cite{klebanov} that the
subleading term is logarithmic in the spin $S$
\begin{equation}
" E-S\sim \ln(S)\ "
\end{equation}
which is essentially the same behavior as found for certain operators on
the gauge theory side
\cite{gross,georgi,floratos, korchemsky, dolan}. The
fluctuations
around the rigidly rotating closed strings were considered in
\cite{tseytlin,tseytlin2} and
confirmed the logarithmic behavior.

The $E$-$N$ relationship has been further investigated in a number of
papers including \cite{russo,armoni,
mandal,minahan,
barbon,axenides,alishabiba,buchel,rashkov,ryang}
for various string configurations. The subleading terms depend on the
particular
string. For instance, for a circular string  it is the square root of $N$
\cite{de
vega, minahan}. It is also known \cite{lomholt,oz} that  torsion may change the
subleading terms.  So this needs further studies.

Another important point is that the $E$-$N$ relation in most cases is
obtained in a purely classical way or from a
simple WKB approximation \cite{de vega,inigo,kar,minahan} of the
path integral, using a method developed in
\cite{hasslacher}.
Quantization of the fluctuations, which is notoriously
a complicated problem for extended strings in curved spacetimes, has only
been done  in very few cases (for instance
the already mentioned \cite{tseytlin,tseytlin2,mandal}).

In the present paper, we consider more generally the fluctuations around
particular
string configurations,  which may be open or closed. Basically, one can
proceed in two
ways: One possibility is to take the Polyakov action in conformal gauge and
to expand
the action and the constraints to second order. However, in a generic curved
spacetime, it is not possible to solve the constraints and thereby to
eliminate the
unphysical modes. The other possibility is to start with the Nambu-Goto
action and
impose more physical gauge conditions and thereby eliminate the unphysical
modes from the very beginning. More
precisely, for a string in a generic $D$-dimensional spacetime, one has,
when using the
Polyakov action in conformal gauge, $D+2$ equations for a complicated
mixture of physical and unphysical
fluctuations. When using the Nambu-Goto action with
suitable
gauge conditions, one has on the other hand only $D-2$ equations for the
physical fluctuations only.
This is why we prefer to use the
Nambu-Goto action supplemented by a term representing the torsion.

The fluctuations around rigidly rotating strings in spacetimes with torsion
were recently considered
in \cite{rashkov,rashkov2}. However, only closed strings were
discussed and the unphysical modes were not eliminated.

The paper is organized as follows:
In section 2, we derive the action, equations of motion and boundary
conditions for quadratic fluctuations around a
classical string configuration in a curved spacetime with torsion. The
fluctuations are split into normal and tangential
contributions, so as to clarify the physical situation. In sections 3 and 4, we
discuss in detail the particular examples
of rigidly rotating open strings in AdS with and without torsion. In
section 5, we comment on the quantization and we
present our conclusions.

\section{General Formalism}
\setcounter{equation}{0}
We are interested in the fluctuations around a classical open planar
string, i.e. a string that is extended only in a 2+1 dimensional section
of a higher-dimensional curved spacetime. It is straightforward to
generalize to completely general configurations, but that seems to be an
unnecessary complication at
the present moment.
As explained in the
introduction, we prefer to work with the Nambu-Goto action with an
additional term representing the torsion.

The action is thus given by
\begin{eqnarray}\label{eq:twodotone}
{\cal S}=\frac{-1}{2\pi\alpha'}\int_M d\tau d\sigma L
\end{eqnarray}
where
\begin{eqnarray}\label{eq:twodottwo}
L=  \sqrt{-\det(
G_{\mu\nu}X^\mu_{\phantom{\mu},\alpha}
X^\nu_{\phantom{\nu},\beta})}-\frac{1}{2}\epsilon^{\alpha\beta}B_{\mu\nu}
X^\mu_{\phantom{\mu},\alpha}X^\nu_{\phantom{\nu},\beta}
\end{eqnarray}
The first variation of the Lagrangian leads to
\begin{eqnarray}
\delta L&=& -\sqrt{-g}\left(g^{\alpha\beta} -
\frac{\epsilon^{\alpha\beta}}{\sqrt{-g}}\right)\hat
K_{\alpha\beta}N_\mu\delta X^\mu \nonumber\\
&&+\partial_\alpha\left[\sqrt{-g}\left(g^{\alpha\beta}G_{\mu\nu}+
\frac{\epsilon^{\alpha\beta}}{\sqrt{-g}}B_{\mu\nu}
\right)X^\mu_{,\beta}\delta X^{\nu}\right]\label{eq:twodotthree}
\end{eqnarray}
where $g_{\alpha\beta}$ is the induced metric on the world-sheet
\begin{equation}
g_{\alpha\beta}=G_{\mu\nu}X^\mu_{\phantom{\mu},\alpha}X^\nu_{\phantom{\nu},\beta}
\end{equation}
while $\hat K_{\alpha\beta}$ is the generalized extrinsic curvature,     as
constructed from the generalized Christoffel-symbols
\begin{eqnarray}
{\hat K}_{\alpha\beta}&=&\left(X^\mu_{\phantom{\mu},\alpha\beta}+
{\hat
\Gamma}^\mu_{\rho\sigma}X^\rho_{\phantom{\rho},\alpha}X^\sigma_{\phantom{\sigma}
,\beta}\right)N_\mu
={\hat
\nabla}_\rho\left(X^\mu_{\phantom{\mu},\alpha}\right)X^\rho_{\phantom{\rho},\beta}N_\mu
\end{eqnarray}
with
\begin{equation}\label{eq:twodotsix}
{\hat
\Gamma}^\mu_{\rho\sigma}=\Gamma^\mu_{\rho\sigma}-\frac{1}{2}H^\mu_{\phantom{\mu}\rho\sigma}
\end{equation}
The equation of motion corresponding to  Eq.(\ref{eq:twodottwo}) is
\begin{equation}\label{eq:twodotseven}
\left(
g^{\alpha\beta}-\frac{\epsilon^{\alpha\beta}}{\sqrt{-g}}\right)\hat
K_{\alpha\beta}=0
\end{equation}
generalizing the usual condition on the extrinsic curvature for a minimal
surface, to
the case of a spacetime with torsion.

For an open string, we get the boundary conditions from the second term in
Eq.(\ref{eq:twodotthree}) (obtained after
projection on the normal and tangential vectors)
\begin{eqnarray}
0&=&[B_{\mu\nu}X^{\mu}_{,\tau}N^\nu]_{\sigma=(0,\pi)}\label{eq:twodoteight}\\
0&=&[\sqrt{-g}-B_{\mu\nu}X^\mu_{,\tau}X^\nu_{,\sigma}]_{\sigma=(0,\pi)}\label{eq:twodotnine}
\end{eqnarray}

Now we turn to the second variation of the Lagrangian.  The derivation
of $\delta^2L$ is a straightforward exercise in differential geometry
following \cite{frolov,guven,guven2,kar2}. But now, keeping all the
surface terms, the result is     ($D_\alpha $ is the world-sheet covariant
derivative)
\begin{align}
L_2&=-\sqrt{-g}\varphi\left[D_\alpha D^\alpha + \hat K_{(\alpha\beta)}
\hat K^{(\alpha\beta)}
-\left(g^{\alpha\beta}-\frac{\epsilon^{\alpha\beta}}{\sqrt{-g}}\right)
\hat R_{\sigma\nu\mu\rho}X^\mu_{,\alpha} X^\nu_{,\beta} N^\rho
N^\sigma\right]\varphi
\nonumber\\
&+\partial_\alpha \left[ \sqrt{-g}\left(2(-g^{\delta \epsilon }\hat
K_{(\delta \gamma)} \varphi +
D_\gamma
\psi^\epsilon)\psi^{[\alpha}\delta^{\gamma]}_\epsilon
+g^{\alpha \beta }(\varphi _{,\beta }+\hat
K_{(\beta\gamma)}\psi^\gamma)\varphi\right)\right]
\nonumber\\
&+ \partial_\alpha \left[ \epsilon^{\alpha\beta}B_{\mu\nu}N^\mu
X^\nu_{,\delta}(\varphi_{,\beta}+\hat
K_{(\beta\epsilon)}\psi^\epsilon)\psi^\delta\right]\nonumber \\
&+\partial_\alpha \left[\epsilon^{\alpha \beta }
B_{\mu\nu} X^\mu_{,\epsilon} (-g^{\epsilon \gamma }\hat
K_{(\gamma \beta)}\varphi+ D_\beta \psi^\epsilon)(N^\nu\varphi +X^\nu_{,\delta
}\psi^\delta )\right]
\nonumber\\
&+ \partial_\alpha\left[
\epsilon^{\alpha\beta}\nabla_\rho B_{\mu\nu}
X^\mu_{,\beta}(N^\rho\varphi+X^\rho_{,\epsilon }\psi^\epsilon
)(N^\nu\varphi+X^\nu_{,\delta }\psi^\delta ) \right]\label{eq:twodotten}
 \end{align}
Here we have expanded the variation of $X^\mu$ on the normal and
tangential vectors
\begin{eqnarray}
\delta X^\mu=\varphi N^\mu + \psi^\alpha X^\mu_{,\alpha}
\end{eqnarray}
such that the normal fluctuations are represented by a world-sheet
scalar field $\varphi$, while the tangential fluctuations are
represented by a world-sheet vector $\psi^\alpha$.

Several comments to
Eq.(\ref{eq:twodotten}) are now in order. First, we notice that only the
symmetric part
of the extrinsic curvature appears. Second, the curvature tensor
$\hat R_{\sigma\mu\nu\rho}$ is the generalized one obtained from the
generalized Christoffel symbols Eq.(\ref{eq:twodotsix}). Third, the
tangential fluctuations
obviously only contribute in the surface terms, i.e. at the boundary,
because of the reparametrization invariance in the bulk. On the other hand, the
surface terms depend explicitly on the torsion $B_{\mu\nu}$ since the
action (\ref{eq:twodotone}) is not invariant under gauge transformations
$\delta
B_{\mu\nu}=\partial_{[\mu}\Lambda_{\nu]}$, but picks up a surface term.

In obtaining the equation (\ref{eq:twodotten}), we also used the first
order equation of
motion (\ref{eq:twodotseven}), but we have not yet used the first order
boundary
conditions (\ref{eq:twodoteight}),(\ref{eq:twodotnine}). The reason for not
having used the first order
boundary conditions at this stage, is that the basis
($N^\mu,X^\mu_{,\alpha}$) generally is not well-defined at the boundary.
Thus, one has to be very careful when implementing the boundary conditions.
This will be clarified in the
following sections.

The equation of motion for the normal fluctuations is then given by
\begin{equation}
\left[D_\alpha D^\alpha + \hat K_{(\alpha\beta)} \hat K^{(\alpha\beta)}
-\left(g^{\alpha\beta}-\frac{\epsilon^{\alpha\beta}}{\sqrt{-g}}\right)\hat
R_{\sigma\nu\mu\rho}X^\mu_{,\alpha} X^\nu_{,\beta} N^\rho
N^\sigma\right]\varphi=0
\end{equation}
while the boundary conditions are
\begin{eqnarray}
0&=&[\sqrt{-g}g^{\sigma\alpha}(\varphi_{,\alpha}+\hat
K_{(\alpha\beta)}\psi^\beta)
+ B_{\mu\nu}N^\mu X^\nu_{,\alpha}(-g^{\alpha \beta }\hat K_{(\beta
\tau)}\varphi+D_\tau\psi^\alpha)
\nonumber \\
&+& \nabla_\nu B_{\mu\rho} N^\mu X^\rho_{,\tau}(N^\nu\varphi
+X^\nu_{,\gamma }\psi^\gamma ) ]_{\sigma=(0,\pi)}
\\
0&=&[-(\sqrt{-g}-B_{\mu\nu}X^\mu_{,\tau}X^\nu_{,\sigma})(-g^{\sigma \alpha
}\hat K_{(\alpha
\tau)}\varphi+D_\tau \psi^\sigma)\nonumber \\
&+&B_{\mu\nu} X^\mu_{,\tau}N^\nu(\varphi_{,\tau}+\hat
K_{(\tau\alpha)}\psi^\alpha)]_{\sigma=(0,\pi)}
\\
0&=&[(\sqrt{-g}-B_{\mu\nu}X^\mu_{,\tau}X^\nu_{,\sigma})(-g^{\tau\alpha}\hat
K_{(\alpha
\tau)}\varphi+D_\tau \psi^\tau)\\
&+&B_{\mu\nu} X^\mu_{,\sigma}N^\nu(\varphi_{,\tau}+\hat
K_{(\tau\alpha)}\psi^\alpha)+\nabla_\nu B_{\mu\rho}
X^\mu_{,\sigma}X^\rho_{,\tau}(N^\nu\varphi+X^\nu_{,\gamma }\psi^\gamma )
]_{\sigma=(0,\pi)}\nonumber
\end{eqnarray}
In the following sections, we shall consider these equations in some particular
cases.

\section{SL(2,R) Background}
\setcounter{equation}{0}
As a first example of our general formalism, we consider the rigidly
rotating open strings in the SL(2,R) $\cong {\rm AdS_3}$ background. One
can for instance imagine a string in the Anti de Sitter part of the
spacetime ${\rm AdS_3}\times {\rm SU(2)}\times {\rm T^4}$. The SL(2,R)
background is given by
\begin{equation}\label{eq:threedotone}
ds^2=-(1+H^2r^2)dt^2+\frac{dr^2}{1+H^2r^2}+r^2d\phi^2,\ \ \
B=-2Hr^2\ dt\wedge d\phi
\end{equation}
There are no straight folded strings in this background
\cite{lomholt,oz}, since the torsion bends and unfolds them. The simplest
open strings, which generalize the straight folded strings in Minkowski
space, are  given by  \cite{lomholt}
\begin{eqnarray}
t&=&c_0\tau \\
r&=&\frac{c_1}{n}\cos(n\sigma ) \\
\phi &=&\frac{c_0\sqrt{n^2+H^2c_1^2}}{c_1}\
\tau+\frac{n^2}{Hc_1}\sigma\nonumber\\
&&-\frac{\sqrt{n^2+H^2c_1^2}}{Hc_1}\cot^{-1} \left(
\frac{\sqrt{n^2+H^2c_1^2}}{n}\cot (n\sigma  )\right)
\end{eqnarray}
where $c_0$ is just introduced for dimensional reasons. It has no
physical importance, so we have a continuous 1-parameter family of solutions
parametrized by $c_1$. The integer $n$ gives the number of string segments
(for $H=0$, the number of foldings).

In the following we take $n=1$, corresponding to the leading Regge
trajectory. Then the
induced metric on the world-sheet is given by
\begin{eqnarray}
g_{\tau\tau}&=&-c_0^2\sin^2\sigma\ ,\\
g_{\tau\sigma}&=&-\frac{Hc_0c_1^2\sqrt{1+H^2c_1^2}\cos^2\sigma\sin^2\sigma}{1+H^
2c_1^2\cos^2\sigma}\ ,\\
g_{\sigma\sigma}&=&\frac{c_1^2\sin^2\sigma
\left[1+\left(2-\cos^2\sigma\right)H^2c_1^2\cos^2\sigma\right]}
{\left(1+H^2c_1^2\cos^2\sigma\right)^2}
\end{eqnarray}
which is singular at the boundary $\sigma=(0,\pi)$. Indeed, the scalar
curvature of the world-sheet is
\begin{equation}
R_2=\frac{2}{c_1^2\sin^4\sigma}
\end{equation}
which is the same as in Minkowski space ($H=0$).

The extrinsic curvature is
\begin{eqnarray}
\hat{K}_{(\tau\tau )}&=&Hc_0^2\cos^2\sigma\ ,\\
\hat{K}_{(\tau\sigma
)}&=&-\frac{c_0\sqrt{1+H^2c_1^2}\left[H^2c_1^2\cos^2\sigma\sin^2\sigma+1\right]}
{1+H^2c_1^2\cos^2\sigma}\ ,\\
\hat{K}_{(\sigma\sigma)}&=&\frac{Hc_1^2\left[2-3\cos^2\sigma+
H^2c_1^2\cos^2\sigma(2\sin^4\sigma-\cos^4\sigma)\right]}
{\left(1+H^2c_1^2\cos^2\sigma\right)^2}\\
{\hat K}_{[\tau\sigma ]}&=&Hc_0c_1\sin^2\sigma\
\end{eqnarray}
 and one can easily verify that
Eqs.(\ref{eq:twodotseven})-(\ref{eq:twodotnine}) are fulfilled.

 The energy and spin angular momentum of these strings are given by
 \cite{lomholt}
\begin{eqnarray}
E&=&\frac{1}{2\pi \alpha'}\int_0^\pi \frac{\partial L}{\partial \dot
t}d\sigma=\frac{c_1}{2\alpha'}\\
S&=&\frac{-1}{2\pi \alpha'}\int_0^\pi \frac{\partial L}{\partial \dot
\phi}d\sigma=\frac{\sqrt{1+H^2c_1^2}-1}{2\alpha'
H^2}\label{eq:threedotfourteen}
\end{eqnarray}
It follows that
\begin{equation}
E=HS\sqrt{1+\frac{1}{H^2\alpha' S}}
\end{equation}
such that for long strings
\begin{equation}\label{eq:threedotsixteen}
E/H-S=\frac{1}{2H^2\alpha'}+\dots
\end{equation}
i.e. the dominant subleading term is just a constant
\cite{lomholt,oz}.

We now turn to the fluctuations. Using that $\hat
R_{\sigma\mu\nu\rho}=0$ for the SL(2,R) background (it is a group manifold),
we get the following
equation of motion for the normal fluctuations
\begin{equation}\label{eq:threedotseventeen}
\left[D^\alpha D_\alpha  + 2H^2 - \frac{2}{c_1^2\sin^4\sigma}\right]\varphi =0
\end{equation}
The boundary conditions  become
\begin{eqnarray}
0&=&\Big[\frac{2H^2c_0c_1(1+H^2c_1^2)\cos^3\sigma}{\sin\sigma(1+H^2c_1^2\cos^2\sigma)
^2}\varphi -\frac{c_0^2\sqrt{1+H^2c_1^2}}{c_1(1+H^2c_1^2\cos^2\sigma)}\psi^\tau
\nonumber\\
&+&\frac{Hc_0c_1[1+2H^2c_1^2\cos^2\sigma
-H^2c_1^2\cos^4\sigma]}{(1+H^2c_1^2\cos^2\sigma)^2}\psi^\sigma
\nonumber\\
&-&\frac{Hc_1\sqrt{1+H^2c_1^2}\cos^2\sigma}{1+H^2c_1^2\cos^2\sigma}\dot\varphi+\frac{c_0}{c_1}\varphi'
+\frac{Hc_0c_1\sin\sigma\cos\sigma}{1+H^2c_1^2\cos^2\sigma}\dot\psi^\tau\label{eq:threedoteighteen}\\
&+&\frac{H^2c_1^3\sqrt{1+H^2c_1^2}\cos^3\sigma\sin\sigma}{(1+H^2c_1^2\cos^2\sigma)^2}\dot\psi^\sigma\Big]_{\sigma=(0,\pi)}
\nonumber\\
0&=&\Big[
-\frac{c_0^2c_1\sin\sigma}{1+H^2c_1^2\cos^2\sigma}
\Big(\frac{\sqrt{1+H^2c_1^2}}{c_1^2\sin\sigma}\varphi+\frac{H\cos\sigma}{c_0}\dot\varphi
+\frac{\sin\sigma}{c_0}\dot\psi^\sigma \nonumber\\
&+&\frac{c_0\cos\sigma (1+H^2c_1^2\cos^2\sigma )}{c_1^2}\psi^\tau
-H\sqrt{1+H^2c_1^2}\cos\sigma \sin^2\sigma \;\psi^\sigma
\Big)\Big]_{\sigma=(0,\pi)}\label{eq:threedotnineteen}\\
0&=&\Big[\frac{Hc_0c_1[1+2H^2c_1^2\cos^2\sigma -H^2c_1^2\cos^4\sigma
]}{(1+H^2c_1^2\cos^2\sigma )^2}\varphi\nonumber\\
&+&\frac{c_0c_1\cos\sigma \sin\sigma [1+2H^2c_1^2\cos^2\sigma
-H^2c_1^2\cos^4\sigma ]}{(1+H^2c_1^2\cos^2\sigma
)^2}\psi^\sigma\nonumber\\
&+&\frac{Hc_0^2c_1\sqrt{1+H^2c_1^2}\cos\sigma \sin^3\sigma
}{1+H^2c_1^2\cos^2\sigma }\psi^\tau
+\frac{c_0c_1\sin^2\sigma }{1+H^2c_1^2\cos^2\sigma }\dot\psi^\tau\nonumber\\
&-&\frac{H^2c_1^3\sqrt{1+H^2c_1^2}\cos^3\sigma \sin\sigma
}{(1+H^2c_1^2\cos^2\sigma )^2}\dot\varphi
\Big]_{\sigma=(0,\pi)}\label{eq:threedottwenty}
\end{eqnarray}
where dot and prime denote derivation with respect to $\tau$ and $\sigma $.
Notice that we keep all the trigonometric functions in the boundary
conditions. The reason is that they have to be expanded, since the
functions ($\varphi, \psi^\alpha$) turn out to be singular at the
boundary, as we will now show.

At this moment we are not interested in the general solution of
Eq.(\ref{eq:threedotseventeen}), which will later be discussed in the
Appendix. Here we only consider the solution near the
boundary, where the boundary
conditions come into play. It is convenient to separate the equations using
\begin{eqnarray}
\varphi(\tau,\sigma)&=&e^{-i\frac{c_0}{c_1}\omega
(\tau+\xi( \sigma))}f_\omega( \sigma  )\label{eq:threedottwentyone}
 \\
\psi^\alpha(\tau,\sigma)&=&e^{-i\frac{c_0}{c_1}\omega
(\tau+\xi( \sigma))}g_\omega^\alpha(\sigma)
\end{eqnarray}
where the function $\xi$ is defined by
\begin{equation}
\xi'=\frac{Hc_1^2\sqrt{1+H^2c_1^2}\cos^2\sigma }{c_0(1+H^2c_1^2\cos^2\sigma )}
\end{equation}
The equation for the normal fluctuations now reduces to
\begin{equation}\label{eq:threedottwentyfour}
f''_\omega+\left(\omega^2+2H^2c_1^2\sin^2\sigma -\frac{2}{\sin^2\sigma }
\right)f_\omega=0
\end{equation}
The solution near the boundary $\sigma=0$ (the analysis near the boundary
$\sigma=\pi$ is completely similar) is
\begin{equation}\label{eq:threedottwentyfive}
f_\omega=\frac{k_1}{\sigma}+k_1\left(
\frac{\omega^2}{2}-\frac{1}{3}\right)\sigma +k_2\sigma^2+\dots
\end{equation}
where $k_1$ and $k_2$ are arbitrary constants. Now
Eqs.(\ref{eq:threedoteighteen})-(\ref{eq:threedottwenty}) lead to
\begin{equation}
g_\omega^\alpha=\frac{d_1^\alpha}{\sigma^2}+\frac{d_2^\alpha}{\sigma}+d_3^\alpha
\end{equation}
where
\begin{eqnarray}
d_1^\tau&=&\frac{-k_1}{c_0\sqrt{1+H^2c_1^2}} \nonumber\\
d_1^\sigma&=&-Hk_1 \nonumber\\
d_2^\tau&=&0 \nonumber\\
d_2^\sigma&=&-i\frac{\omega
k_1\sqrt{1+H^2c_1^2}}{c_1}\label{eq:threedottwentyseven}
\end{eqnarray}
while the finite terms are related by
\begin{eqnarray}
0&=&\frac{k_1}{6}\left( -2 -18H^2c_1^2 +3\omega^2 -10H^4c_1^4 +
6H^2c_1^2\omega^2 + 3H^4c_1^4\omega^2\right) \nonumber\\
&-& c_0(1+H^2c_1^2)^{3/2}d_3^\tau +
Hc_1^2(1+H^2c_1^2)d_3^\sigma
\end{eqnarray}
It follows that both the normal and the tangential  fluctuations  are
infinite at the
boundary. However, if we consider the physical  fluctuations,  which at the
boundary $\sigma =0$, are given by
\begin{equation}\label{eq:threedottwentynine}
\delta X^\mu(\sigma =0)=[\varphi N^\mu +\psi^\alpha X^\mu_{,\alpha}
]_{\sigma =0}
\end{equation}
we find that the singularities precisely cancel, and we get a finite result
\begin{align}
\delta
t(\sigma =0)=&\left(
\frac{k_1[3\omega^2(1+H^2c_1^2)+2H^2c_1^2-4]}{6(1+H^2c_1^2)^{3/2}}+d_3^\tau
c_0\right)e^{-i\frac{c_0}{c_1}\omega(\tau+\xi(0))}\\
\delta r(\sigma =0)=&\,i\omega
k_1\sqrt{1+H^2c_1^2}e^{-i\frac{c_0}{c_1}\omega(\tau+\xi(0))}\\
\delta
\phi(\sigma
=0)=&{\Bigg(\frac{k_1[3\omega^2(1+H^2c_1^2)+8H^2c_1^2+2]}{6c_1(1+H^2c_1^2)}}\nonumber\\
&\quad\quad\quad\quad\quad\quad\quad\quad{+\frac{c_0\sqrt{1+H^2c_1^2}}{c_1}d_3^\tau\Bigg)e^{-i\frac{c_0}{c_1}\omega(\tau+\xi(0))}}
\end{align}
It should be stressed that one can always obtain finite normal and tangential
fluctuations at one
  of the boundaries by choosing appropriate values of the
integration constants. For instance, taking
  $k_1=0$ in
Eqs.(\ref{eq:threedottwentyfive})-(\ref{eq:threedottwentyseven}) makes
everything finite at $\sigma =0$.
However, one can easily show (by considering
  the exact solution in some simple case, say $H=0$; see the Appendix)
  that both normal and tangential fluctuations then still blow up
  at the boundary $\sigma =\pi$. What we have shown here is that the
physical fluctuations in any case are
  finite everywhere.

\section{Ordinary Anti de Sitter Space}
\setcounter{equation}{0}
As a second example, we now consider the fluctuations around the
rigidly rotating open string in ordinary anti de Sitter space.
Again, we take 2+1 dimensions, which might be envisioned as a
slice of ${\rm AdS_5}$ in  ${\rm AdS_5}\times {\rm S^5}$.

This solution was originally constructed in \cite{inigo}. The radial
coordinate  is
usually expressed in terms of an elliptic function, but for
comparison with section 3, we now use the following gauge
\begin{equation}\label{eq:fourdotone}
t=c_0\tau, \ \ \ r=\frac{c_1}{n}\cos(n\sigma  )
\end{equation}
The remaining coordinate is obtained by the ansatz $\phi=\omega\tau$,
where $\omega$ is determined by the boundary conditions. Actually,
this is the solution for a rigidly rotating open string (rotating
around its center of mass) in an arbitrary static cylindrically
symmetric spacetime with line element
\begin{equation}
ds^2=g_{tt}(r)dt^2+g_{rr}(r)dr^2+g_{\phi\phi}(r)d\phi^2+2g_{t\phi}(r)dt
d\phi
\end{equation}
and without torsion,
as one can easily verify since the induced metric is diagonal while
the extrinsic curvature has only the $\tau\sigma$-component. In the
case of Anti de Sitter space, where the line element is given in
Eq.(\ref{eq:threedotone}), one finds that
\begin{equation}\label{eq:fourdotthree}
\phi=\frac{c_0\sqrt{n^2+H^2c_1^2}}{c_1}\tau
\end{equation}
which is exactly the same time dependency as in the case considered in
section 3.       It must be stressed, though, that the solution considered
here is not a special case of the solution
considered in section 3. The two solutions only overlap for $H=0$, i.e. in
Minkowski space.

In the present case, the induced metric on the world sheet becomes (again
we take the leading Regge
trajectory, $n=1$)
\begin{eqnarray}
g_{\tau\tau}=-c_0^2\sin^2\sigma ,\ \ \
g_{\sigma\sigma}=\frac{c_1^2\sin^2\sigma }{1+H^2c_1^2\cos^2\sigma }
\end{eqnarray}
which is again singular at the boundary $\sigma=(0,\pi)$, as follows
from the scalar curvature
\begin{equation}
R_2=
\frac{-2}{c_1^2\sin^4\sigma }[H^2c_1^2\cos^4\sigma
-2H^2c_1^2\cos^2\sigma-1]
\end{equation}
In the absence of torsion, the extrinsic curvature is just the usual
one (symmetric in its indices), which we now call $K_{\alpha\beta}$.
In the present case it is given by
\begin{eqnarray}
{K}_{\tau\tau}&=&K_{\sigma \sigma }=0 \nonumber\\
K_{\tau \sigma }&=& -c_0
\frac{\sqrt{1+H^2c_1^2}}{\sqrt{1+H^2c_1^2\cos^2\sigma}}
\end{eqnarray}
and trivially $g^{\alpha\beta}K_{\alpha\beta}=0$, as mentioned above.

In this case, the energy and spin are given by \cite{inigo}
\begin{eqnarray}
 E&=&\frac{c_1\sqrt{1+H^2c_1^2}}{\pi\alpha'
}{\rm E}\left(\frac{H^2c_1^2}{1+H^2c_1^2}\right)\label{eq:fourdotseven}\\
 S&=&\frac{1+H^2c_1^2}{\pi\alpha'H^2}\left[
     {\rm E}\left( \frac{H^2c_1^2}{1+H^2c_1^2}\right)
     -\frac{1}{1+H^2c_1^2} {\rm
K}\left(\frac{H^2c_1^2}{1+H^2c_1^2}\right)
\right]
\end{eqnarray}
It is impossible to express $E$ directly in terms of $S$, but for long
strings where the modulus
of the elliptic integrals goes towards $1$, one
finds
\cite{klebanov}
\begin{equation}\label{eq:fourdotnine}
E/H-S=\frac{1}{2\pi \alpha' H^2}\ln(2\alpha' H^2 S)+\dots
\end{equation}
which should be compared with Eq.(\ref{eq:threedotsixteen}). Notice that
the   factors  of $2$ are not present
 for the corresponding closed folded string ($n=2$).

Eqs.(\ref{eq:fourdotseven})-(\ref{eq:fourdotnine}) are well-known, but we
now turn to the
fluctuations.  The fluctuations were actually already considered   in some
cases in \cite{kar},  but without imposing any
boundary conditions for the fluctuations.
For the corresponding folded closed string,  where there are of course no
boundary conditions at all, the fluctuations were considered in
\cite{tseytlin,tseytlin2}.

The equation of motion for the normal fluctuations is
\begin{equation}\label{eq:fourdotten}
\left[D^\alpha D_\alpha-2H^2-2\frac{1+H^2c_1^2}{c_1^2\sin^4\sigma
}\right]\varphi=0
\end{equation}
while the boundary conditions are
\begin{eqnarray}
0&=&\left[\varphi'-\frac{c_0\sqrt{1+H^2c_1^2}}{\sqrt{1+H^2c_1^2\cos^2\sigma}}
\psi^\tau\right]_{\sigma=(0,\pi)}\label{eq:fourdoteleven}\\
0&=&[c_0\sqrt{1+H^2c_1^2}\sqrt{1+H^2c_1^2\cos^2\sigma}\
\varphi+c_1^2\sin^2\sigma\dot\psi^\sigma
\nonumber\\
&& +c_0^2\cos\sigma\sin\sigma(1+H^2c_1^2\cos^2\sigma)\psi^\tau
]_{\sigma=(0,\pi)}\\
0&=&\left[\sin^2\sigma\dot\psi^\tau+\cos\sigma\sin\sigma\psi^\sigma
\right]_{\sigma=(0,\pi)}\label{eq:fourdotthirteen}
\end{eqnarray}
First we write
\begin{equation}
\varphi(\tau,\sigma)=    e^{-i\frac{c_0}{c_1}\omega\tau}f_\omega(\sigma),\ \ \
\psi^\alpha(\tau,\sigma)=e^{-i\frac{c_0}{c_1}\omega\tau}g_\omega^\alpha(\sigma )
\end{equation}
Now Eq.(\ref{eq:fourdotten}) becomes
\begin{eqnarray}
0&=&(1+H^2c_1^2\cos^2\sigma
)\frac{d^2f_\omega}{d\sigma^2}-H^2c_1^2\cos\sigma \sin\sigma
\frac{df_\omega}{d\sigma}\nonumber \\ &+&
\left( \omega^2
-2\frac{1+H^2c_1^2}{\sin^2\sigma}-2
H^2c_1^2\sin^2\sigma\right) f_\omega\label{eq:heun}
\end{eqnarray}
This equation is actually well-known in the mathematical literature where
it is known
as ``Heun's equation"; see the Appendix.  Again we only consider the
solution near the
boundary $\sigma=0$. The solution is
\begin{equation}
f_\omega=\frac{k_1}{\sigma}+\frac{-k_1}{6(1+H^2c_1^2)}\left(5H^2c_1^2+
2-3\omega^2\right)\sigma+k_2\sigma^2+\dots
\end{equation}
while the boundary conditions
(\ref{eq:fourdoteleven})-(\ref{eq:fourdotthirteen}) lead to
\begin{equation}
g^\alpha_\omega=\frac{d^\alpha_1}{\sigma^2}+\frac{d^\alpha_2}{\sigma}
+d^\alpha_3
\end{equation}
where
\begin{eqnarray}
d^\tau_1&=&-\frac{k_1}{c_0}\nonumber  \\
d^\tau_2&=&0\nonumber \\
d^\tau_3&=&-\frac{k_1}{6c_0(1+H^2c_1^2
)}\left(2H^2c_1^2+2-3\omega^2\right)\nonumber \\
d^\sigma_1&=&0\nonumber \\
d^\sigma_2&=&-\frac{i\omega k_1}{c_1}
\end{eqnarray}
while $d^\sigma_3$ is arbitrary. We notice that both the normal and
tangential fluctuations are infinite at the boundary. However, the
physical fluctuations  given by Eq.(\ref{eq:threedottwentynine})
turn out to be finite  also in this case
\begin{eqnarray}
\delta t(\sigma=0)
&=&-\frac{k_1}{1+H^2c_1^2}\left(1+H^2c_1^2-\omega^2\right)e^{-i\frac{c_0}{c_1}
\omega\tau}\\
\delta r (\sigma=0)&=&i\omega k_1e^{-i\frac{c_0}{c_1}
\omega\tau}\\
\delta\phi
(\sigma=0)&=&\frac{k_1}{c_1\sqrt{1+H^2c_1^2}}\left(-H^2c_1^2+\omega^2\right)e^{-i\frac{c_0}{c_1}
\omega\tau}
\end{eqnarray}
with similar conclusions as at the end of section 3.

If we take $n=2$ in Eqs.(\ref{eq:fourdotone}),(\ref{eq:fourdotthree}) the
result can be interpreted as a
folded closed  string.
This is of course the philosophy in \cite{klebanov,tseytlin}.     In that
case, the tangential
fluctuations completely decouple and we are left with the normal
fluctuations which diverge
at the boundaries. Our arguments therefore seem to hold only for open
strings. This is a problem
which deserves further study.

\section{Concluding Remarks}
\setcounter{equation}{0}
 We derived the equations of motion for normal fluctuations around a
classical   string in a curved spacetime
 with torsion. The boundary conditions, involving both normal and
tangential fluctuations, for open strings were also
obtained.  In the two cases of rigidly rotating open strings in AdS with
and without torsion, it was shown that
the divergent part of the tangential fluctuations at the boundary is
completely
determined  by the divergent part of the
normal fluctuations in such a way that the physical fluctuations are
everywhere finite. This is a non-trivial result
since the classical  world-sheet is singular at the boundary.  It should be
stressed that this does not work for
the closed folded strings (which exist in the absence of torsion), since
the tangential fluctuations decouple
completely  in that case.

The relation between normal and tangential fluctuations at the boundary
also shows that one cannot in advance set the
tangential fluctuations equal to zero everywhere, since this would kill
also the normal fluctuations.

The quantization of the normal fluctuations is a difficult task. It is
equivalent to the quantization of a scalar field
in a curved spacetime. Thus, it is necessary  to make some approximations.
          In the cases of rigidly rotating
strings,  the problem is that the mass-term is $\sigma$-dependent.  It was
argued in \cite{tseytlin} that the mass-term
can be approximated by a constant, though, at least for long strings.
Moreover, earlier arguments
\cite{tseytlin3,tseytlin4} suggest that at the quantum level
one should actually skip completely the part of the mass-term
corresponding to $R_2$. In any case, one then ends up with the quantization
of an ordinary massive scalar field.

The general formalism developed in section 2 simplifies considerably
for closed strings, where the
tangential fluctuations decouple completely and there are no boundary
conditions.
The closed  strings, recently considered in connection with the AdS/CFT
duality, on AdS$_5\times$S$^5$
and AdS$_3\times$S$^3\times$T$^4$ (for instance \cite{russo,oz} and
references given therein),
can then be straightforwardly analyzed. This is currently under investigation.

\vskip 24pt
\hspace*{-6mm}{\bf Acknowledgements}:\\
We would like to thank  N. Sibani for his help in the preparation of this
paper.

\appendix
\section{Appendix}
\setcounter{equation}{0}
In this appendix we consider the equations for the fluctuations
(\ref{eq:heun}) in
more detail. Equation (\ref{eq:heun}) is
\begin{eqnarray}
0&=&(1+H^2c_1^2\cos^2\sigma
)\frac{d^2f_\omega}{d\sigma^2}-H^2c_1^2\cos\sigma \sin\sigma
\frac{df_\omega}{d\sigma}\nonumber \\ &+&
\left( \omega^2
-2\frac{1+H^2c_1^2}{\sin^2\sigma}-2
H^2c_1^2\sin^2\sigma\right) f_\omega\label{eq:adotone}
\end{eqnarray}
First, we introduce the conformal coordinate $\tilde\sigma$
\begin{equation}
\cos\sigma={\rm cn}\left[ \sqrt{1+H^2c_1^2}\ \tilde\sigma,\ m\right]
\end{equation}
where cn is the Jacobi elliptic function with modulus
$m=H^2c_1^2/(1+H^2c_1^2)$. Then equation (\ref{eq:adotone})
becomes
                \begin{equation}
\frac{d^2f_\omega}{d\tilde\sigma^2} +\left( \omega ^2
-2\frac{(1+H^2c_1^2)}{{\rm sn}^2}-2
H^2c_1^2{\rm sn}^2
\right) f_\omega=0\label{eq:adotthree}
        \end{equation}
with ${\rm sn}={\rm sn}\left[ \sqrt{1+H^2c_1^2}\ \tilde\sigma,\ m\right]$.
Notice that (\ref{eq:adotthree}) generalizes the Lam\'{e} equation in the
same way as
eq.(\ref{eq:threedottwentyfour})
generalizes the P\"{o}schl-Teller equation \cite{flugge}. However, we can
go one
step further by introducing the coordinate
\begin{equation}
z=\sin^2\sigma ={\rm sn}^2\left[ \sqrt{1+H^2c_1^2}\ \tilde\sigma,\ m\right]
\end{equation}
Then eq.(\ref{eq:adotthree}) reduces to
        \begin{eqnarray}
0&=&\frac{d^2g}{dz^2}+\left[\frac{5/2}{z}+\frac{1/2}{z-1}+\frac{1/2
}{z-(1/H^2c_1^2+1)}\right]\frac{dg}{dz }
\nonumber \\
&+& \frac{z+(\omega ^2/4H^2c_1^2
-1/H^2c_1^2-2)}{[z-(1/H^2c_1^2+1)]z(z-1)} g\label{eq:adotfive}
\end{eqnarray}
where $g=f_\omega/z$. Eq.(\ref{eq:adotfive}) is a special case of Heun's
equation
\cite{poole}
\begin{eqnarray}
0&=&\frac{d^2g}{dz^2}+\left[\frac{\gamma}{z}+\frac{\delta}{z-1}+\frac{\epsilon
}{z-a}\right]\frac{dg}{dz }  +\frac{\alpha\beta z-q}{(z-a)z(z-1)} g
\end{eqnarray}
where $\alpha+\beta +1=\gamma +\delta +\epsilon $. The parameters are
$\alpha =2,\ \beta =1/2,\
\gamma =5/2,\ \delta =1/2,\ \epsilon =1/2$.
The accessory parameters are $q=-(\omega^2/4H^2c_1^2-1/H^2c_1^2-2)$ and
$a=1/H^2c_1^2+1$.

For $H=0$ the solution to eq.(\ref{eq:adotfive}) is just a hypergeometric
function. For
general $H$ the solution can be expanded on hypergeometric functions, but we
shall not go into the details here.


\newpage
\begin{thebibliography}{99}
\bibitem{sanchez} A.L. Larsen and N.
S\'{a}nchez, Phys. Rev. {\bf D52}, 1051
(1995).
\bibitem{de vega} H.J. de Vega, A.L.
Larsen and N. S\'{a}nchez, Phys. Rev. {\bf
D51}, 6917 (1995).
\bibitem{inigo} H.J. de Vega and  I.L.
Equsquiza, Phys. Rev. {\bf D54}, 7513 (1996).
\bibitem{maldacena} J. Maldacena,
Adv. Math. Phys. {\bf 2}, 231 (1998),
Int. J. Theor. Phys {\bf 38}, 1113 (1999).
\bibitem{gubser} S.S. Gubser, I.R.
Klebanov and  A.M. Polyakov, Phys. Lett. {\bf
B428}, 105 (1998).
\bibitem{witten} E. Witten,
Adv. Theor. Math. Phys. {\bf 2}, 253 (1998).
\bibitem{klebanov} S.S. Gubser, I.R.
Klebanov and A.M. Polyakov, Nucl. Phys. {\bf
B636}, 99 (2002).
\bibitem{gross} D.J. Gross and F. Wilczek,
Phys. Rev. {\bf D9}, 980 (1974).
\bibitem{georgi} H. Georgi and D. Politzer, Phys. Rev. {\bf D9},
416 (1974).
\bibitem{floratos} E.G. Floratos, D.A. Ross and  C.T. Sachrajda,
Nucl. Phys. {\bf B129}, 66 (1977).
\bibitem{korchemsky} G.P. Korchemsky, Mod. Phys. Lett. {\bf
A4}, 1257 (1989).
\bibitem{dolan} F.A. Dolan and H. Osborn, Nucl. Phys. {\bf
B629}, 3 (2002).
\bibitem{tseytlin} S. Frolov and  A.A.
Tseytlin, JHEP {\bf 0206}, 007 (2002).
\bibitem{tseytlin2} A.A. Tseytlin,
Int. J. Mod. Phys {\bf A18}, 981 (2003).
\bibitem{russo} J.G. Russo, JHEP {\bf
0206}, 038 (2002).
\bibitem{armoni} A. Armoni, J.L.F. Barbon and
A.C. Petkou, JHEP {\bf 0206}, 058 (2002).
\bibitem{mandal} G. Mandal, Phys. Lett.
{\bf B543}, 81 (2002).
\bibitem{minahan} J.A. Minahan, Nucl. Phys.
{\bf B648}, 203 (2003).
\bibitem{barbon} A. Armoni, J.L.F. Barbon and
A.C. Petkou, JHEP {\bf 0210}, 069 (2002).
\bibitem{axenides} M. Axenides and  E.
Floratos, ``Scaling Violations in
Yang-Mills Theories and Strings in
ADS(5)", hep-th/0210091 (unpublished).
\bibitem{alishabiba} M. Alishahiha and  A.E.
Mosaffa, JHEP {\bf 0210}, 060 (2002).
\bibitem{buchel} A. Buchel, Phys. Rev. {\bf
D67}, 066004 (2003).
\bibitem{rashkov} R.C. Rashkov and  K.S.
Viswanathan, ``Rotating Strings with
B-Field", hep-th/0211197 (unpublished).
\bibitem{ryang} S. Ryang, ``Rotating and
Orbiting Strings in the Near Horizon Brane
Backgrounds", hep-th/0303237 (unpublished).
\bibitem{lomholt} M.A. Lomholt and  A.L.
Larsen, Int. J. Mod. Phys {\bf A18}, 1007
(2003).
\bibitem{oz} A. Loewy and  Y. Oz, Phys. Lett.
{\bf B557}, 253 (2003).
\bibitem{kar} S. Kar and  S. Mahapatra,
Class. Quant. Grav. {\bf 15}, 1421 (1998).
\bibitem{hasslacher} R. Dashen, B. Hasslacher and A. Neveu, Phys. Rev. {\bf
D11},
3424 (1975).
\bibitem{rashkov2} H. Dimov, V. Filev,
R.C. Rashkov and K.S. Viswanathan,
``Semiclassical Quantization of Rotating
Strings in Pilch-Warner Geometry",
hep-th/0304035 (unpublished).
\bibitem{frolov} A.L. Larsen and  V.P. Frolov,
Nucl. Phys. {\bf B414}, 129 (1994).
\bibitem{guven} J. Guven, Phys. Rev. {\bf D48}, 4604 (1993).
\bibitem{guven2} J. Guven, Phys. Rev. {\bf D48}, 5562 (1993).
\bibitem{kar2} S. Kar, Phys. Rev. {\bf
D54}, 6408 (1996).
\bibitem{tseytlin3} N. Drukker, D.J. Gross and A.A. Tseytlin, JHEP {\bf
0004}, 021
(2000).
\bibitem{tseytlin4} A.A. Tseytlin, `` 'Long' Quantum
Superstrings in ADS(5) $\times$ S$^5$", hep-th/0008107
(unpublished).
\bibitem{flugge} S. Fl\"{u}gge, Practical Quantum Mechanics I (Springer-Verlag,
New York, 1971).
\bibitem{poole} E.G.C. Poole, Introduction to the Theory of Linear Differential
Equations (Clarendon Press, Oxford, 1936).
\end{thebibliography}
\end{document}